\documentclass[prl,twocolumn,amsmath,amssymb]{revtex4}

\usepackage{ulem}
\usepackage{graphicx}% Include figure files
\usepackage{dcolumn}% Align table columns on decimal point
\usepackage{bm}% bold math
\usepackage{xcolor}

\begin{document}

\title{Collisionless Sound in a Uniform Two-Dimensional Bose Gas}

\author{Miki Ota$^1$, Fabrizio Larcher$^{1,2}$, Franco Dalfovo$^1$, Lev Pitaevskii$^{1,3}$, Nikolaos P. Proukakis$^2$, and Sandro Stringari$^1$}

\affiliation{$^1$ INO-CNR BEC Center and Dipartimento di Fisica, Universit\`a di Trento, 38123 Trento, Italy}
\affiliation{$^2$ Joint Quantum Centre Durham--Newcastle, School of Mathematics, Statistics and Physics, Newcastle University, Newcastle upon Tyne, NE1 7RU, United Kingdom}
\affiliation{$^3$ Kapitza Institute for Physical Problems RAS, Kosygina 2, 119334 Moscow, Russia}

\date{\today}

\begin{abstract}
Using linear response theory within the random phase approximation, we investigate the propagation of sound in a uniform  two dimensional (2D) Bose gas in the collisionless regime. We show that the sudden removal of a static density perturbation produces a damped oscillatory behavior revealing that sound can propagate also in the absence of collisions, due to mean-field interaction effects. We provide explicit results for the sound velocity and damping as a function of temperature, pointing out the crucial role played by Landau damping.  We support our predictions by performing numerical simulations with the stochastic (projected) Gross-Pitaevskii equation. The results are consistent with the recent experimental observation of sound in a weakly interacting 2D Bose gas both below and above the superfluid Berezinskii-Kosterlitz-Thouless transition.
\end{abstract}

\maketitle

In classical hydrodynamics, sound is a density wave that propagates due to collisions between particles. In superfluids, the situation is more complex. If collisions are strong enough to ensure local thermalization, Landau's  two-fluid hydrodynamics predicts the existence of two sounds, first and second sound \cite{Landau1941,Landau1987,Khalatnikov,Griffin2007,Griffin2009,Verney2015,Pitaevskii2017}, the latter being very sensitive to the value of the superfluid density. If temperature is low enough ($T\ll T_c$), density waves propagate due to coherence and interaction and, in a weakly interacting Bose gas, they take the form of Bogoliubov sound \cite{book}.  Density waves have been observed in harmonically trapped 3D Bose gases at $T\sim 0$ \cite{Andrews1997} and at finite temperature \cite{Meppelink2009}.  Very recently, measurements of sound propagation have become available also in a uniform 2D Bose gas \cite{Dalibard2018}. Such a system is of particular interest since, in two dimensions, the velocity of second sound is predicted to vanish with a finite jump at the Berezinskii-Kosterlitz-Thouless (BKT) phase transition \cite{Ozawa2014}. In fact, while in three dimensions the superfluid density of a dilute Bose gas can be directly related to the condensate fraction \cite{book, Pethick2002}, in two dimensions  it remains finite even if Bose-Einstein condensation is  ruled out by the Hohenberg-Mermin-Wagner theorem \cite{Hohenberg1967, Mermin1966}. The BKT phase transition is of infinite order \cite{Berezinskii1972, Kosterlitz1973, Hadzibabic2011} and does not show any discontinuity in the thermodynamic quantities \cite{Yefsah2011}, but the superfluid density exhibits a universal jump, with a consequent discontinuity of the speed of second sound \cite{Ozawa2014}. However, the experiment of Ref. \cite{Dalibard2018} does not reveal the occurrence of any jump in the sound velocity, whose value is found to remain finite above $T_c$ and significantly smaller than the one expected for the first (adiabatic) sound. 

The key issue for understanding these observations is the role of collisions, which would be essential for the application of two-fluid hydrodynamics.  In the quasi-2D regime of Ref. \cite{Dalibard2018}, where the frequency of the transverse harmonic confinement is such that $\hbar \omega_z \gg k_B T$, the collisional rate is given by $\Gamma^\mathrm{coll} =  \hbar n \tilde{g}^2/ m$ \cite{Petrov2001}, where $n$ is the 2D density, $\tilde{g} = m g_{2D}/\hbar^2 = \sqrt{8\pi} a/ \ell_z$ is the dimensionless coupling constant, $a_s$ is the $s$-wave scattering length, and $l_z=\sqrt{\hbar/m\omega_z}$. In Ref. \cite{Dalibard2018}, the collisional rate is of the same order of the frequency of the excited mode, determined by the box length, and this clearly suggests that collisions are not efficient enough to ensure the collisional hydrodynamic regime. Hence one needs a theory that can describe density waves in the absence of collisions and, above $T_c$, even in the absence of superfluidity. 

An appropriate starting point consists of combining the Boltzmann transport equation with linear response theory. In fact, from a self-consistent formulation of the Boltzmann  equation in the absence of the collisional term, one can derive the random phase approximation (RPA) expression \cite{Griffin2009, SM} 
\begin{equation}\label{Eq.1}
\chi (k,\omega) = \frac{\chi_0(k,\omega)}{1+g_{2D} \chi_0(k,\omega)} \, , 
\end{equation}
for the dynamic response function per particle of a 2D Bose gas, where $\chi_0(k,\omega)$ is the response function of a noninteracting Bose gas. At low temperature, RPA is known to be equivalent to Bogoliubov theory. In particular, at $T \to 0$, one has $\chi_0(k,\omega) \sim (-n \hbar^2 k^2/m)/\left[(\hbar\omega)^2 - (\hbar^2 k^2/2m)^2 \right]$,  and the poles of $\chi(k,\omega)$ coincide with the Bogoliubov dispersion relation $\hbar \omega = \sqrt{\hbar^2 k^2 g_{2D}n/m + (\hbar^2k^2/2m)^2}$. In the same regime, the formalism is well suited to investigate the thermodynamic functions at equilibrium. At finite temperature, Eq.~(\ref{Eq.1}) accounts for Landau damping and for the role of interactions at the mean-field level. It is worth mentioning that, in three dimensions and at temperatures higher than $T_c$, the value of the coupling constant should be multiplied by $2$ as a result of exchange effects (see Sect.13.2 in Ref. \cite{book}); however, in a weakly interacting 2D Bose gas, these effects are expected to remain small even at $T \gtrsim T_c$ because density fluctuations tend to be suppressed by the persistence of a degenerate quasicondensate, provided $T$ remains smaller than the degeneracy temperature $T^*=2\pi \hbar^2 n_{2D}/(mk_B)$ \cite{SM, Prokofev2001,Prokofev2002,Bisset2009} and, consequently, we do not include the factor $2$ in our analysis.  Even though RPA ignores quantum fluctuations associated with the critical region near $T_c$, it can nevertheless serve as a first description of the dynamic behavior of the gas in the absence of collisions. 

%%%%%%%%%%%%%%%%%%%%%%%%%%%%%%%%%%%%%%%%%%%%%%%%%%%%%%%%%%%%%%%%%%%%%%%%%%
\begin{figure}[t]
\begin{center}
\includegraphics[width=0.9\columnwidth]{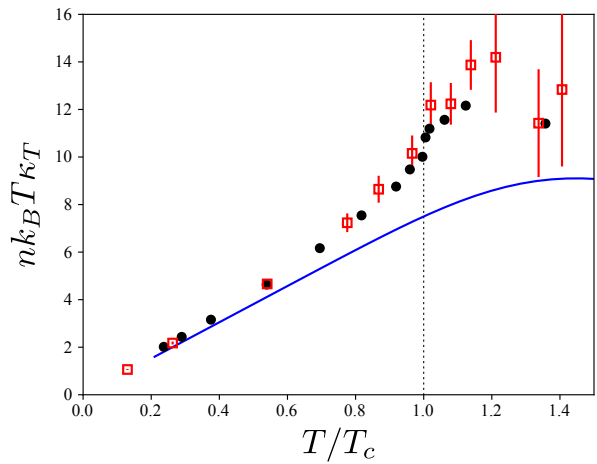}
\caption{Dimensionless isothermal compressibility as a function of temperature for $g_{2D}=0.1\hbar^2/m$. The blue solid line and the red squares correspond to $\kappa_T$ calculated with the RPA expression \eqref{Eq.3} and with SGPE, respectively. The error bars of the SGPE data represent the statistical uncertainty of the average over different noise realizations. The black dots are the results obtained from the universal relations of Refs.~\cite{Prokofev2001,Prokofev2002}.} 
\label{fig:Fig_kappaT}
\end{center}
\end{figure}
%%%%%%%%%%%%%%%%%%%%%%%%%%%%%%%%%%%%%%%%%%%%%%%%%%%%%%%%%%%%%%%%%%%%%%%%%%

In the long wavelength (small $k$) limit the response function $\chi_0(k,\omega)$ of the ideal Bose gas takes the simplified form 
\begin{equation}
\label{Eq.2}
\chi_0(k,\omega) =  \int \frac{\mathrm{d}^2 \mathbf{p}}{(2\pi\hbar)^2}  \frac{\partial f_0}{\partial p_x} \  \frac{1}{\omega / k - p_x/m  + i\delta} \, ,
\end{equation} 
with $\delta \rightarrow 0^+$, where $f_0(\textbf{p})=[e^{(p^2/(2m)-\mu^\mathrm{IBG})/(k_BT)}-1]^{-1}$ is the Bose distribution function of the 2D ideal Bose gas and the chemical potential $\mu^\mathrm{IBG}$ is  fixed  by the normalization condition $n = (2\pi\hbar^2)^{-2}\int \mathrm{d}^2\textbf{p} f_0(\textbf{p})$. At small $k$, the response function only depends on the dimensionless velocity $u= (\omega/k)\sqrt{m/(2k_BT)}$. Expression \eqref{Eq.1} satisfies the $f$-sum rule $\chi (k, \omega \rightarrow \infty) =  -n k^2/(m \omega^2)$, and the long wavelength limit of the static  polarizability $\chi (k \rightarrow 0 , \omega = 0) = n^2  \kappa_T$ \cite{Nozieres} (compressibility sum rule) leads to the following result for the isothermal compressibility of the gas:
\begin{equation}\label{Eq.3}
\kappa_T = \frac{m}{2\pi\hbar^2n^2} \frac{1}{\left[ e^{-\mu^\mathrm{IBG}/(k_BT)}-1\right] +g_{2D} m/(2 \pi \hbar^2)} \, .
\end{equation}
The  isothermal compressibility $\kappa_T$ is expected to play an important role in characterizing  the dynamic behavior of the gas in the collisionless regime, differently from the adiabatic compressibility $\kappa_S$ which instead describes the propagation of sound in the collisional regime. In particular, using these two quantities one can define the isothermal sound velocity $c_T= \sqrt{1/(mn\kappa_T)}$ and the adiabatic sound velocity $c_S=\sqrt{1/(mn\kappa_S)}$. It is thus useful to check the quality of  the RPA prediction for $\kappa_T$ by comparing it with other many-body approaches. In Fig.~1 we show the prediction of Eq.~(3) (solid line), together with the theoretical results obtained from the universal relations of Refs.~\cite{Prokofev2001, Prokofev2002, Rancon2012} (black circles). For the value $g_{2D}=0.1\hbar^2/m$ used in the figure, the critical temperature, defined as in Ref. \cite{Prokofev2001},  is $T_c= 0.12T^*$. The RPA curve is in qualitative agreement with the universal relations, but underestimates the maximum near the critical temperature, where quantum fluctuations, neglected in RPA,  have significant effects.

We also calculate $\kappa_T$ from the equilibrium solutions of a stochastic (projected) Gross-Pitaevskii equation (SGPE). In this approach, all states of the system lying below a certain energy cutoff  are described by a classical function $\Psi$ \cite{Stoof,Blakie} obeying a Gross-Pitaevskii equation where dissipation and stochastic fluctuations are included. The energy cutoff is chosen in such a way that the mean occupation number of states below the cutoff is larger than $1$. The equilibrium at a given $T$ is ensured by coupling $\Psi$ with the incoherent (sparsely occupied) states above the cutoff, acting as a thermal bath. Given the stochastic nature of the model, the physical observables are obtained by averaging over many noise realizations. Using this theory, we calculate the equation of state for equilibrium configurations of a 2D Bose gas at different temperatures and we use it to extract the isothermal compressibility \cite{SM}.  As shown in Fig.~\ref{fig:Fig_kappaT}, though SGPE includes fluctuations within an approximate scheme, it reproduces the peak of $\kappa_T$ near $T_c$ in quantitative agreement with the universal relations of Refs. \cite{Prokofev2001,Prokofev2002}. 

%%%%%%%%%%%%%%%%%%%%%%%%%%%%%%%%%%%%%%%%%%%%%%%%%%%%%%%%%%%%%%%%%%%%%%%%%%
\begin{figure}[t]
\begin{center}
\includegraphics[width=0.8\columnwidth]{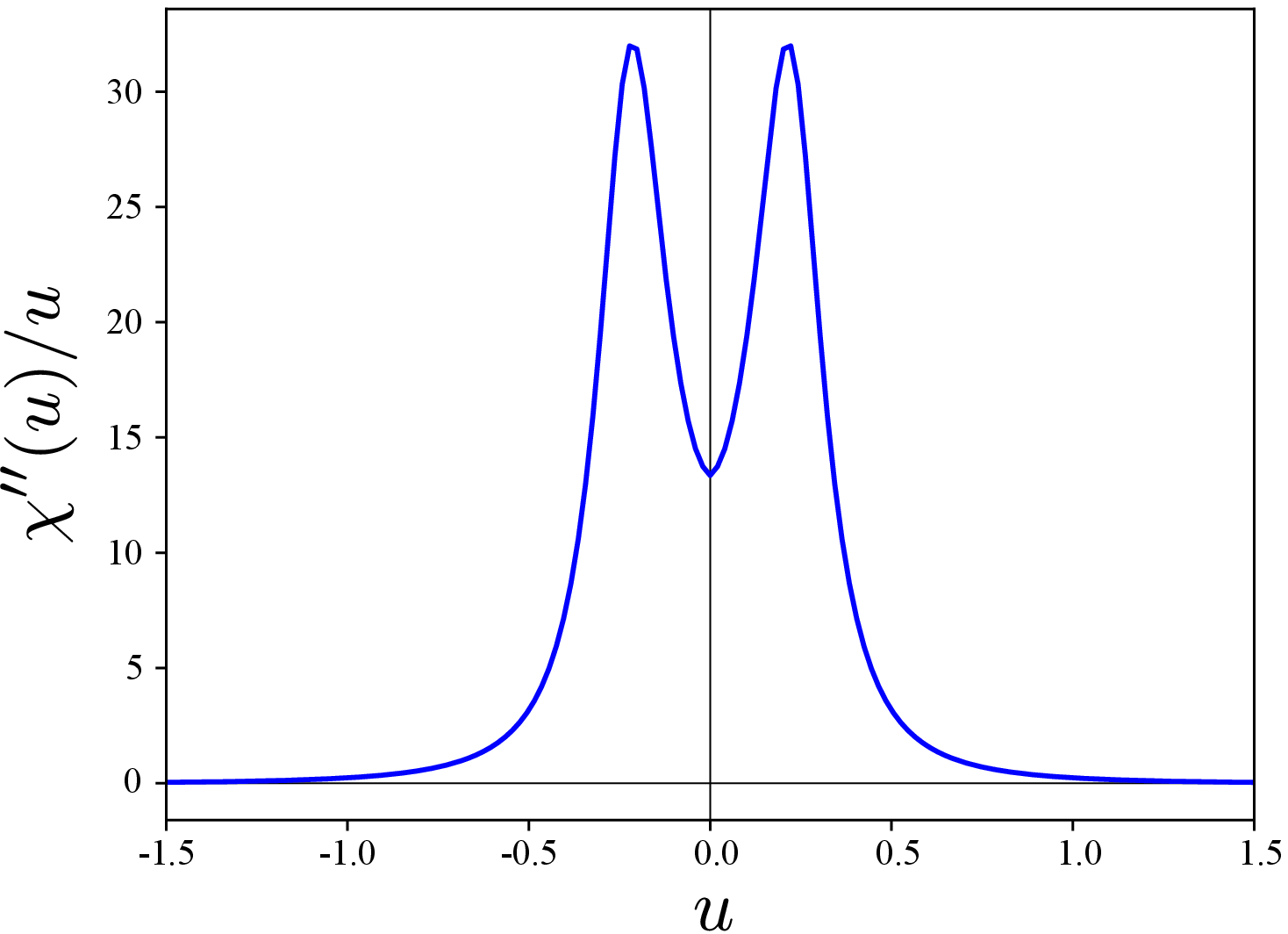}\\
\includegraphics[width=0.8\columnwidth]{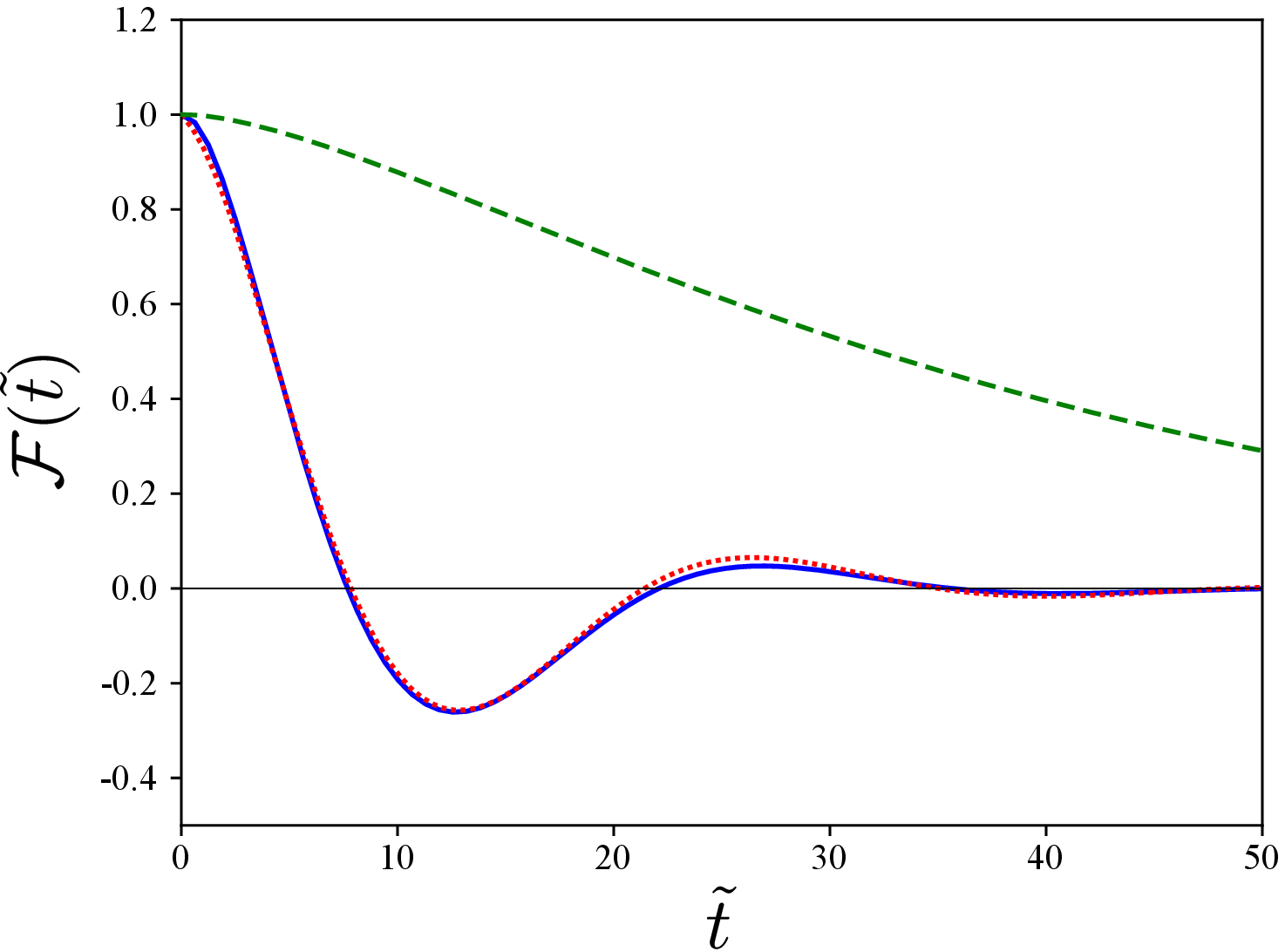}
\caption{Upper panel: Imaginary part of the inverse frequency weighted response function, $\chi''(u)/u$, calculated in the RPA at $T=0.15T^*=1.2T_c$, with $g_{2D}=0.16\hbar^2/m$, as a function of the dimensionless frequency $u= (\omega/k)\sqrt{m/(2k_BT)}$. Lower panel: Fourier transform \eqref{Eq.5}, as a function of the dimensionless time $\tilde{t}=k t \sqrt{2k_BT/m}$. The blue solid and green dashed lines correspond to the interacting and ideal gas, respectively. The red dotted line is the fit based on Eq.~\eqref{Eq.6}.} 
\label{fig:Fig_Chi}
\end{center}
\end{figure}
%%%%%%%%%%%%%%%%%%%%%%%%%%%%%%%%%%%%%%%%%%%%%%%%%%%%%%%%%%%%%%%%%%%%%%%%%%

\begin{figure}[t]
\begin{center}
\includegraphics[width=0.9\columnwidth]{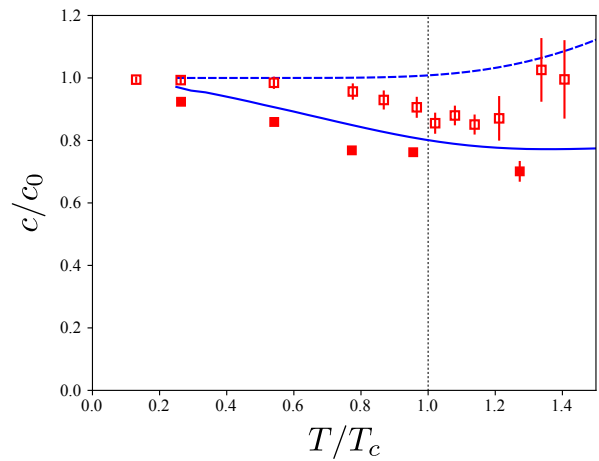}
\caption{Sound velocity in units of $c_0=\sqrt{g_{2D}n/m}$ for $g_{2D}=0.1\hbar^2/m$. The blue solid line is the sound velocity $c={ \tilde \omega}/k$ extracted from the Fourier transform Eq.~\eqref{Eq.5} of $\chi''/\omega$ calculated in RPA, while the blue dashed line is the isothermal sound velocity $c_T= \sqrt{1/(mn\kappa_T)}$, within the same theory. Filled and empty squares represent the sound speed extracted from real time simulations and the isothermal sound velocity, respectively, both obtained with SGPE. The error bars arise from statistical fluctuations; below $T_c$ they are of the same order of the marker size.}
\label{fig:Fig_c}
\end{center}
\end{figure}

In order to calculate the speed and damping of density waves from the RPA expression (\ref{Eq.1}), inspired by the experimental procedure, we first consider the gas at equilibrium in the presence of a weak, spatially periodic, stationary potential, producing a sinusoidal density modulation with a given wave vector $k$. Then, if the external potential is suddenly removed, the density starts oscillating with a time dependent amplitude given by \cite{Zambelli2001, Arahata2009} 
\begin{equation}\label{Eq.5}
\mathcal{F} (t) = \frac{1}{\pi n^2\kappa_T} \int_{-\infty}^\infty \mathrm{d}\omega \frac{\chi '' (k,\omega)}{\omega} e^{i\omega t} \, , 
\end{equation}
where the signal is normalized to its $t=0$ initial value, fixed by the isothermal compressibility.  If the ratio $\chi '' (k,\omega)/\omega$ exhibits a narrow peak, as happens at low temperature, then the oscillation will persist for a long time; if instead the same function is broad, then the oscillation is strongly damped. Hence the function  $\mathcal{F}(t)$  provides direct information on the velocity of sound and on its damping. 
In the upper panel of Fig.~\ref{fig:Fig_Chi} we show a typical profile of the  function $\chi '' (k,\omega)/\omega$ calculated from Eq.~\eqref{Eq.1} with $g_{2D}=0.16\hbar^2/m$ at $T=1.2T_c$. The figure reveals the occurrence of a peak at $\omega \ne 0$, which is at the origin of a damped oscillatory behavior in the Fourier transform $\mathcal{F} (t)$,  shown in the lower panel. By  using 
\begin{equation}\label{Eq.6}
\mathcal{F}^{DHO} (t) = e^{-\Gamma t/2} \left( \cos(\tilde{\omega} t) + \frac{\Gamma}{2\tilde{\omega}} \sin(\tilde{\omega}t) \right),
\end{equation}
as a fitting function \cite{NoteDHO} we can estimate the velocity of sound, $c={ \tilde \omega}/k$, and its damping rate $\Gamma$ \cite{note-poles}. The resulting velocity $c$ is shown in Fig.~\ref{fig:Fig_c} (solid line) as a function of $T$, in units of the zero temperature Bogoliubov sound velocity $c_0=\sqrt{g_{2D}n/m}$; the curve is close to the prediction for the isothermal sound velocity  $c_T$ determined by the isothermal compressibility (dashed line). The adiabatic sound velocity $c_S$, which describes the propagation of sound in the collisional regime, is not shown in the figure and lies well above $c_T$ ($c_S/c_T \sim 2$ near $T_c$). It is worth noticing on passing that the oscillatory behavior of the function $\mathcal{F} (t)$ is caused by the interaction term in the denominator of $\chi(k,\omega)$. In fact, in the ideal Bose gas ($g_{2D}=0$), the function $\chi '' (k,\omega)/\omega$ has a peak at $\omega=0$ and its Fourier transform \eqref{Eq.5} is a monotonically decreasing function (dashed line in Fig.~\ref{fig:Fig_Chi}). 

%%%%%%%%%%%%%%%%%%%%%%%%%%%%%%%%%%%%%%%%%%%%%%%%%%%%%%%%%%%%%%%%%%%%%%%%%
%%%%%%%%%%%%%%%%%%%%%%%%%%%%%%%%%%%%%%%%%%%%%%%%%%%%%%%%%%%%%%%%%%%%%%%%%%

The propagation of density waves at finite temperature can also be numerically simulated with SGPE. To this purpose, consistent with the RPA calculations, we suddenly remove an external static sinusoidal perturbation and let the gas evolve in time. Simulations are performed in a rectangular box with periodic boundary conditions. The wave vector of the excited mode is fixed by the box size by $k=2\pi n_x/L_x$, and we typically use $L_x= 40$ and $n_x=1,2,3$.  The initial state is an equilibrium configuration prepared  with SGPE at a given $T$ and then the real time evolution is performed with a (projected) GPE for the classical field only, by removing the coupling with the incoherent states above the cutoff \cite{Blakie,Davis,Proukakis}, and averaging over many realizations. We then extract the values of $c$ and $\Gamma$ by using expression \eqref{Eq.6} as a fitting function for the amplitude of the observed density oscillations. The same speed $c$ is found, within the statistical uncertainties, from a linear fit to the mode frequency $\tilde{\omega}(k)$ obtained in simulations with different $n_x$. The results are shown in Fig.~\ref{fig:Fig_c} as solid red squares, while the open squares represent the isothermal sound velocity  $c_T$, with $\kappa_T$ taken from equilibrium solutions of SGPE. The figure shows that both RPA and SGPE predict a sound velocity which remains roughly constant near $T_c$ and reasonably close to the isothermal velocity $c_T$; the discrepancy between $c_T$ calculated with RPA (dashed line) and SGPE (empty squares) is consistent with the difference between the corresponding values of $\kappa_T$ in Fig.~\ref{fig:Fig_kappaT}. 

%%%%%%%%%%%%%%%%%%%%%%%%%%%%%%%%%%%%%%%%%%%%%%%%%%%%%%%%%%%%%%%%%%%%%%%%%%
\begin{figure}[t]
\begin{center}
\includegraphics[width=0.9\columnwidth]{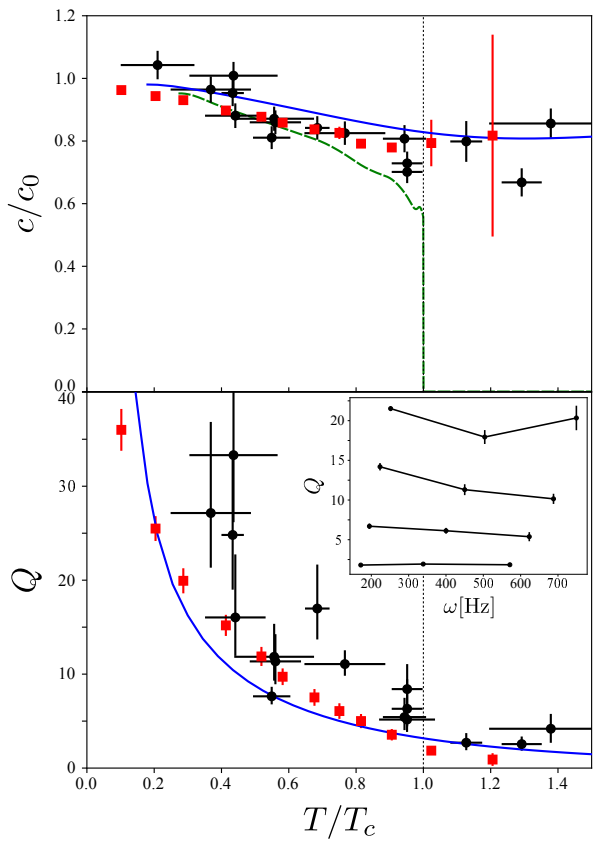}
\caption{Upper panel shows the sound velocity calculated for $g_{2D}=0.16 \hbar^2/m$. Black circles: experimental data of Ref.~\cite{Dalibard2018}; blue solid line:  RPA; red squares: SGPE; green dashed line: second sound predicted by Landau's two-fluids hydrodynamics \cite{Ozawa2014, Ota2018}.  Lower panel shows the quality factor $Q=2\tilde{\omega} / \Gamma$. The blue solid line is $Q$ evaluated from the Fourier transform $\mathcal{F}$ in RPA; black circles are experimental data \cite{Dalibard2018}; red squares are the results of SGPE simulations, obtained by averaging over simulations done at different values of excited frequency (see inset). The inset shows $Q$ as a function of frequency at different values of $T$ from SGPE; from top to bottom: $T/T_c=0.29,~0.52,~0.75,~1.02$. The error bars of the SGPE data in both panels represent the statistical deviations due to different noise realizations.}
\label{fig:Fig_expt}
\end{center}
\end{figure}
%%%%%%%%%%%%%%%%%%%%%%%%%%%%%%%%%%%%%%%%%%%%%%%%%%%%%%%%%%%%%%%%%%%%%%%%%%

In  Fig.~\ref{fig:Fig_expt} we compare our results with the experimental observations of Ref.~\cite{Dalibard2018}. In the upper panel we show the sound velocity calculated using RPA  (solid line) and SGPE (red squares), while black circles are the experimental results. Theory and experiments  reasonably agree both below and above  $T_c$.  Below $T_c$, our predictions for the velocity of the collisionless sound are close to the ones for second sound based on Landau's two-fluid hydrodynamics (dashed green line). This is not surprising since, for a weakly interacting Bose gas at temperatures larger than $g_{2D}n_{2D}/k_B$, the velocity of second sound is well approximated by the expression $\sqrt{(n_s/n)/(mn\kappa_T)}$ \cite{Ozawa2014} and differs from the isothermal velocity only by the multiplicative factor $\sqrt{n_s/n}$, which is fixed by the superfluid fraction and is $\sim 0.7$ near $T_c$. Conversely, neither our theoretical sound velocity nor the experimental one exhibit the jump to zero at $T_c$, which would be predicted by two-fluids hydrodynamics  in the collisional regime \cite{Ozawa2014, Ota2018}. 

The lower panel shows the quality factor $Q=2\tilde{\omega} / \Gamma$. By increasing the temperature, $Q$ decreases as the damping rate becomes quickly large. Above $T_c$,  damping becomes so strong that the oscillatory behavior is hardly visible. Again, there is an overall good agreement between theory and experiments. In RPA, the behavior of $Q$ is the consequence of Landau damping, i.e., the coupling between the collective sound oscillation and the (thermally populated) single-particle excited states included in the ideal Bose gas response (\ref{Eq.2}) (see also Ref. \cite{Chung2009} for similar results). In SGPE,  the same mechanism is accounted for by the dynamical coupling between excited states described by the classical field below the cutoff energy. This is confirmed by the independence of $Q$ on frequency, as shown in the inset of Fig.~\ref{fig:Fig_expt}; in fact, if damping were collisional, it would exhibit a quadratic increase with $\omega$ and hence a pronounced frequency dependence of the quality factor. 

In conclusion, our theoretical predictions, based on the random phase approximation and on the stochastic (projected) Gross-Pitaesvkii equation,  strongly support the interpretation of the recent experimental results of \cite{Dalibard2018} in terms of the propagation of sound in the collisionless regime. The signatures of this sound have been explored by looking at the time evolution of the system after the sudden removal of a spatially periodic  perturbation.  Our work reveals that collisionless sound can propagate  both below and above the BKT transition, as a consequence of the interaction between particles.    

An issue that remains to be explored concerns the development of a theory for second sound which includes superfluid effects also in a  collisionless regime (see, for example, \cite{Khalatnikov}). Another important issue is the  crossover between collisionless and collisional regimes which may be relevant for more strongly interacting 2D Bose gases, or in larger boxes. An increase of the collisional rate by a factor of $10$ is likely within  the present experimental reach \cite{Ha2013}. By moving towards the collisional regime, one expects to observe a decrease of damping accompanied by an increase of the sound velocity above $T_c$, where it should approach the adiabatic sound velocity $c_S$. One also expects to observe the jump of second sound at $T_c$ and its conversion, above the transition, into a diffusive mode, which would significantly contribute to the isothermal compressibility sum rule \cite{Nozieres}. These studies would complement the investigation of sound propagation in strongly interacting Fermi gases \cite{Sidorenkov2013, Zwierlein2018}, where collisional hydrodynamics is expected to dominate.

\acknowledgements

{\bf Acknowledgements}. We thank the authors of Ref. \cite{Dalibard2018} for stimulating discussions and for providing us with their experimental data. We also thank S. ~Giorgini and  T.~Ozawa for many useful comments. This project has received funding from the EU Horizon 2020 research and innovation programme under Grant Agreement No. 641122 QUIC, and by Provincia Autonoma di Trento.

\setcounter{figure}{0}   
\setcounter{equation}{0}   
\renewcommand{\thefigure}{S\arabic{figure}}
\renewcommand{\theequation}{S\arabic{equation}}

\section{Supplemental Material}

This Supplemental Material is meant to provide some additional information about the theory presented in the paper, in the form of a brief summary of known facts about the Random Phase Approximation and the Stochastic Gross-Pitaevskii equations, and technical details about the calculations. 

\section{S1. Mean-field approximation}

Equation (1) of the main paper gives the dynamic response function of the 2D Bose gas in the Random Phase Approximation (RPA). It is consistent with the mean-field expression
\begin{equation}\label{Eq.A1}
P = \frac{1}{2} g_{2D}n^2 + \int \frac{d^2\mathbf{p}}{(2 \pi \hbar)^2} \frac{p^2}{2m} f_0 (\mathbf{p})  ,
\end{equation}
for the pressure of the interacting 2D Bose gas, where $f_0 ({\bf p})= \{\exp [(p^2/2m-\mu^\mathrm{IBG})/k_B T]-1\}^{-1}$ is the single particle distribution function and  $\mu^\mathrm{IBG}$ is the chemical potential fixed by the normalization condition $n=(2\pi \hbar)^{-2}\int d^2{\bf p }f_0({\bf p})$. Starting from the above expression, one can derive the thermodynamic functions of the gas.  In particular, one obtains the isothermal compressibility $\kappa_T$ given in Eq.~(3) of the main paper by calculating the derivative $n \kappa_T = (\partial P/ \partial n)^{-1}$ at fixed $T$.
\par
The interaction term $g_{2D} n^2/2$ entering the expression for $P$ differs by the factor $1/2$ from the corresponding term predicted by Hartree-Fock theory in 3D above the BEC critical temperature \cite{book}. In 3D, the exchange term in the thermal component is responsible for doubling the value of the density fluctuations with respect to the $T=0$ case, yielding an extra factor $2$ in the interaction energy. Conversely, in 2D, density fluctuations are significantly reduced also above the Berezinskii-Kosterlitz-Thouless transition due to the persistence of a quasi-condensation regime  \cite{Prokofev2001, Prokofev2002}. Using a suitable definition of condensate order parameter within SGPE, we numerically verified that the gas remains in a quasi-condensate regime in the  temperature range relevant to our work, including $T$ above $T_c$, but still smaller than the quantum degeneracy temperature $T^*=2\pi \hbar^2n_{2D}/(mk_B)$, in agreement with previous predictions based on similar classical field methods \cite{Bisset2009}. In this range, density fluctuations are strongly suppressed and we are justified to use the $T=0$ expression for the interaction energy in $P$. 
\par
The formalism can be easily generalized to include the long wavelength variations of the distribution function induced by an external potential  $U\left(\mathbf{r},t\right) =U_0 \exp(ikx-i\omega t)$. The kinetic equation for the (space and time dependent) distribution function  $f({\bf r}, {\bf p}, t)$ takes the form of the Boltzmann transport equation without collisional term. In the limit of small perturbations, one gets \cite{Kadanoff}:
\begin{equation}\label{Eq.A3}
i\left( kv_{x}-\omega \right) \delta f-\frac{\partial f_{0}}{\partial p_{x}} g_{2D}ik\int \frac{\mathrm{d}^{2}\mathbf{p}}{\left( 2\pi \hbar \right) ^{2}} \delta f-\frac{\partial f_{0}}{\partial p_{x}}ikU_0=0 \; ,
\end{equation}
where $f (\mathbf{p},\mathbf{r},t) = f_0 (\textbf{p}) + \delta f (\textbf{p}) \exp [i(kx-\omega t)]$, 
and $f_0(\textbf{p})$ is the equilibrium distribution. From this equation, one can directly calculate the density response function $\chi (k,\omega)$  defined by the relation \cite{book, Hu2010}
\begin{equation}
\delta n=\int \frac{\mathrm{d}^{2}\mathbf{p}}{\left( 2\pi \hbar \right) ^{2}} \delta f = -U_0 \chi (k,\omega),
\end{equation}%
from which one finally derives the RPA expression used in the paper,
\begin{equation}\label{Eq.A4}
\chi (k,\omega) = \frac{\chi_0(k,\omega)}{1+g_{2D} \chi_0(k,\omega)} \, .
\end{equation}
The knowledge of the RPA response function \eqref{Eq.A4} permits to identify the poles in the complex plane, allowing for the determination of the real and imaginary parts of the eigenfrequencies. At temperatures satisfying the conditions $T \ll T^*$ and  $(\omega / k)^2 \ll k_BT/m$, one can use the expansion $f_0(E) \simeq k_BT/E$ and calculate $\chi_0$ analytically. This yields the following equation for the poles:
\begin{equation}\label{Eq.A5}
1 + \tilde{g} \left\lbrace \frac{1}{2 \pi u^2} \left[ 1 - \ln (2 u) + \frac{1}{2} \ln \eta \right] +  \frac{i}{4} \frac{u}{(u^2+\eta)^{3/2}} \right\rbrace = 0,
\end{equation}
with $u= (\omega/k)\sqrt{m/(2k_BT)}$ and $\eta=- \mu^\mathrm{IBG}/k_BT$. The explicit solution of Eq. \eqref{Eq.A5} confirms the results discussed in the main text for the real and imaginary parts of the collective frequencies.

\section{S2. Stochastic Gross-Pitaevskii Equation}

In numerical simulations, we treat the system by means of the stochastic (projected) Gross-Pitaevskii equation (SGPE) \cite{Stoof,Blakie}  \begin{equation}\label{eq:SGPE}
	i\hbar \frac{\partial \Psi}{\partial t} = \hat{P} \left\{ \left(1- i \gamma \right)\left[-\frac{\hbar^2 \nabla^2}{2m} + U + g|\Psi|^2 -\mu \right]\Psi + \eta \right\}
\end{equation}
where $\Psi({\bf r},t)$ is a classical field, $U({\bf r})$ is the external potential,  $g$ is the coupling constant of the mean-field interaction,  $\mu$ is the chemical potential, $\gamma$ is a parameter fixing the strength of dissipation, and $\eta({\bf r},t)$ is a stochastic term with white noise correlation
\begin{equation}
	\langle \eta^*({\bf r},t)\eta({\bf r}',t')\rangle = 2 \gamma  \hbar  k_\text{B} T\ \delta(t-t') \delta({\bf r}-{\bf r}')  \, .
\end{equation}
Differently from the case the usual Gross-Pitaevskii equation at $T=0$, where $\Psi$ is the macroscopic wave function representing the condensate, the complex function $\Psi$ in SGPE does not only represent the condensate, but it also contains a finite number of excited states above it, up to an energy cutoff. In our calculations, the cutoff is typically fixed at the value $\mu+k_BT \ln 2$, which corresponds to a mean occupation number of order $1$. The function $\Psi$ is coupled {\it via} $\gamma$ and $\eta$ to the reservoir of classical thermal atoms, corresponding to the sparsely occupied high-energy states above the cutoff. The value of $\gamma$ is usually fixed to reproduce typical experimental growth rates (i.e., the rate at which the spontaneous accumulation of atoms in the low-lying modes occurs until the equilibrium is reached). The projector $\hat{P}$ ensures that only states below the cutoff are included in the classical field during the system evolution. 

In our paper, we have considered a gas tightly confined in the transverse direction, so that the SGPE is solved in 2D with an effective coupling constant $g_{2D}$, defined in the same way as in the RPA theory above. The parameters in the simulations are those of a gas of $^{87}$Rb atoms. For the size of the computational box we used $L_x \times L_y = (50 \times 50) \mu$m for the data in Fig.~1 and $(40 \times 30) \mu$m for the data in Figs.~3 and 4. The number of atoms is typically around $30000$.

Equilibrium configurations have been obtained by evolving the system from pure noise both for a uniform gas ($U=0$) and with a sinusoidal external potential $U({\bf r})=U_0 \sin (kx)$, with an integer number of wavelengths in the box, and using  periodic boundary conditions in both cases. The uniform configurations are used to calculate the equation of state (including atoms both below and above cutoff) and the isothermal compressibility. The latter is obtained by first evaluating the phase space density $n\lambda_{\rm dB}^2$, where $\lambda_{\rm dB}= h/\sqrt{2\pi m k_BT}$ is the thermal de Broglie wavelength, and then taking its derivative with respect to the variable $\mu/T$. Since points at different $T$ are subject to statistical fluctuations and the derivative is calculated with the finite difference method, the resulting values of $\kappa_T$ in Fig.~1 of the paper are slightly scattered, especially above $T_c$.   

The configurations with the initial external potential are instead used to produce collective modes. We suddenly remove the external potential and, at the same time we remove the terms in $\gamma$ and $\eta$ from the SGPE, so that the system evolves in time obeying a projected Gross-Pitaevskii equation for the classical field, stochastically generated at $t=0$ \cite{Blakie,Davis,Proukakis}. From the evolution we extract the speed of sound and the damping rate as explained in the paper. We have checked that, for $U_0 \sim 0.1 \mu$ or smaller, the response to the initial perturbation is linear and thus the results become independent of $U_0$. Sound speed and damping rates are calculated for two values of the coupling constant: $g_{2D}=0.1$ and $0.16 \hbar^2/m$.   

Simulations are run on Newcastle University's High-Performance-Computing cluster, Topsy, with the software package XMDS2 \cite{xmds}. The overall CPU time is approximately 30000 hours, with an average of about $20$ nodes working in parallel.

%%%%%%%%%%%%%%%%%%%%%%%%%%%%%%%%%%%%%%%%%%%%%%%%%%%%%%%%%%%%%%%%%%%%%%%%%%

% Bibliography

%%%%%%%%%%%%%%%%%%%%%%%%%%%%%%%%%%%%%%%%%%%%%%%%%%%%%%%%%%%%%%%%%%%%%%%%%%


\begin{thebibliography}{99}

\bibitem{Landau1941} L. D. Landau, J. Phys. USSR \textbf{5}, 71 (1941).
\bibitem{Landau1987} L. D. Landau and E. M. Lifshitz, \textit{Fluid Mechanics} (Pergamon, Oxford, 1987).
\bibitem{Khalatnikov} I.M. Khalatnikov, \textit{An Introduction to the Theory of Superfluidity} (W. A. Benjamin, New York, 1965).
\bibitem{Griffin2007} A. Griffin and E. Zaremba, Phys. Rev. A {\bf 56}, 4839 (1997).
\bibitem{Griffin2009} A. Griffin, T. Nikuni, and E. Zaremba, {\it Bose-Einstein Condensed Gases at Finite Temperatures} (Cambridge University Press, Cambridge, 2009).
\bibitem{Verney2015} L. Verney, L. P. Pitaevskii, and S. Stringari, Europhys. Lett. \textbf{111}, 40005 (2015).
\bibitem{Pitaevskii2017} L. P. Pitaevskii and S. Stringari, in \textit{Universal Themes of Bose-Einstein Condensation}, edited by  N. P. Proukakis, D. W. Snoke, and P. B. Littlewood (Cambridge University Press, Cambridge, 2017), pp. 322-347.
\bibitem{book}  L. Pitaevskii and S. Stringari, {\it Bose-Einstein Condensation and Superfluidity} (Oxford University Press, 2016)
\bibitem{Andrews1997} M. R. Andrews, D. M. Kurn, H.-J. Miesner, D. S. Durfee, C. G. Townsend, S. Inouye,  and W. Ketterle,  Phys. Rev. Lett. {\bf 79}, 553 (1997).
\bibitem{Meppelink2009} R. Meppelink, S. B. Koller, and P. van der Straten, Phys. Rev. A \textbf{80}, 043605 (2009).
\bibitem{Dalibard2018} J. L. Ville, R. Saint-Jalm, E. Le Cerf, M. Aidelsburger, S. Nascimb{\`{e}}ne, J. Dalibard, and J. Beugnon, arXiv:1804.04037 (2018).
\bibitem{Ozawa2014} T. Ozawa and S. Stringari, Phys. Rev. Lett. \textbf{112}, 025302 (2014).
\bibitem{Pethick2002} C. J. Pethick and H. Smith, {\it Bose-Einstein Condensation in Dilute Gases} (Cambridge University Press, Cambridge, 2002).
\bibitem{Hohenberg1967} P. C. Hohenberg, Phys. Rev. \textbf{158}, 383 (1967).
\bibitem{Mermin1966} N. D. Mermin and H. Wagner, Phys. Rev. Lett. \textbf{17}, 1133 (1966).
\bibitem{Berezinskii1972} V. L. Berezinskii, Sov. Phys. JETP \textbf{34}, 610 (1972).
\bibitem{Kosterlitz1973} J. M. Kosterlitz and D. J. Thouless, J. Phys. C \textbf{6}, 1181 (1973).
\bibitem{Hadzibabic2011} Z. Hadzibabic and J. Dalibard, Riv. Nuovo Cimento \textbf{34}, 389 (2011).
\bibitem{Yefsah2011} T. Yefsah, R. Desbuquois, L. Chomaz, K. J. G{\"{u}}nter, and J. Dalibard, Phys. Rev. Lett. \textbf{107}, 130401 (2011).
\bibitem{Petrov2001} D. S. Petrov and G. V. Shlyapnikov, Phys. Rev. A \textbf{64}, 012706 (2001).
\bibitem{SM} see Supplemental Material for more details.
\bibitem{Prokofev2001} N. Prokof'ev, O. Ruebenacker, and B. Svistunov, Phys. Rev. Lett. \textbf{87}, 270402 (2001).
\bibitem{Prokofev2002} N. Prokof'ev and B. Svistunov, Phys. Rev. A \textbf{66}, 043608 (2002).
\bibitem{Bisset2009} R. N. Bisset, M. J. Davis, T. P. Simula, and P. B. Blakie, Phys. Rev. A {\bf 79}, 033626 (2009); R.N. Bisset and P.B. Blakie, Phys. Rev. A {\bf 80}, 045603 (2009).
\bibitem{Nozieres} P. Nozi\`eres and D. Pines, \textit{The Theory of Quantum Liquids} (Addison-Wesley, Redwood City, 1989), Vol. II.
\bibitem{Rancon2012} A. Ran{\c{c}}on and N. Dupuis, Phys. Rev. A \textbf{85}, 063607 (2012).
\bibitem{Stoof} H. T. C. Stoof and M. J. Bijlsma, J. Low Temp. Phys. {\bf 124}, 431 (2001).
\bibitem{Blakie} P.B. Blakie, A.S. Bradley, M.J. Davis, R.J. Ballagh, and C.W. Gardiner, Adv. Phys. {\bf 57}, 363 (2008).
\bibitem{Zambelli2001} F. Zambelli and S. Stringari,  Phys. Rev. A \textbf{63}, 033602 (2001).
\bibitem{Arahata2009} E. Arahata and T. Nikuni, Phys. Rev. A \textbf{80}, 043613 (2009).
\bibitem{NoteDHO} The choice of this fitting function is suggested by the behavior of the Fourier transform $\mathcal{F}(t)$ derived from the response function $\chi^\mathrm{DHO}(\omega) = (\omega_0^2 -\omega^2 -i\omega\Gamma)^{-1}$ of the damped harmonic oscillator, where $\omega_0$ and $\Gamma$ are the frequency and damping rate of the oscillator, respectively.  
\bibitem{note-poles} The frequencies and damping rates can also be calculated from the poles of the response function (\ref{Eq.1}), as the real and imaginary parts of the eigenfrequencies \cite{SM}. We have verified that, in the relevant temperature range, including the region near $T_c$, the two procedures give indistinguishable results.
\bibitem{Davis} M. J. Davis, S. A. Morgan, and K. Burnett, Phys. Rev. Lett. {\bf 87}, 160402 (2001).
\bibitem{Proukakis} N. P. Proukakis, J. Schmiedmayer, and H. T. C. Stoof, Phys. Rev. A {\bf 73}, 053603 (2006).
\bibitem{Ota2018} M. Ota and S. Stringari, Phys. Rev. A \textbf{97}, 033604 (2018).
\bibitem{Chung2009} M.-C. Chung and A. B. Bhattacherjee, New J. Phys. \textbf{11}, 123012 (2009).
\bibitem{Ha2013} L.-C. Ha, C.-L. Hung, X. Zhang, U. Eismann, S.-K. Tung, and C. Chin, Phys. Rev. Lett. \textbf{110}, 145302 (2013).
\bibitem{Sidorenkov2013} L. A. Sidorenkov, M. K. Tey, R. Grimm, Y.-H. Hou, L. P. Pitaevskii, and S. Stringari, Nature \textbf{498}, 78 (2013).
\bibitem{Zwierlein2018} M. W. Zwierlein, talk at the \textit{BEC 2017 Frontiers in Quantum Gases}, Sant Feliu de Guixols, September 2 - 8, 2017.
\bibitem{Kadanoff} L. P. Kadanoff and G. Baym, \textit{Quantum Statistical Mechanics} (W.A. Benjamin, New York, 1962).
\bibitem{Hu2010} H. Hu, E. Taylor, X.-J. Liu, S. Stringari, and G. Griffin, New J. Phys. \textbf{12}, 043040 (2010).
\bibitem{xmds} G.R.~Dennis, J.J. Hope, and M.T. Johnsson, Comput. Phys. Commun. {\bf 184}, 201 (2013). 

\end{thebibliography}
\end{document}